\documentclass[conference]{IEEEtran}
\usepackage[utf8]{inputenc}
\usepackage{booktabs}
\usepackage{subcaption}
\usepackage{algorithm}
\usepackage{algpseudocode}
\usepackage{tikz}
\usetikzlibrary{arrows.meta,positioning,shapes,calc}
\usepackage{pgfplots}
\pgfplotsset{compat=1.18}
\usepackage{cite}
\usepackage{amsmath,amssymb,amsfonts}
\usepackage{graphicx}
\usepackage[inkscapelatex=false]{svg}
\usepackage{tabularx}
\usepackage{textcomp}
\usepackage{xcolor}
\usepackage{multirow} 
\usepackage{array}  
\usepackage{float} 
\usepackage{hyperref}
\usepackage{xcolor}
\usepackage{orcidlink}
\hypersetup{
    colorlinks=true,
    linkcolor=blue,
    citecolor=blue,
    filecolor=magenta,      
    urlcolor=cyan,
    pdftitle={},
    pdfpagemode=FullScreen,
}

\begin{document}


\title{A Four-Tier Communication Architecture and Sim-to-Real Validation of a Graphical Open-Source Platform for Robotic Engineering Education}



\author{\IEEEauthorblockN{
    Thien Tran\IEEEauthorrefmark{1},
    Khang Duong\IEEEauthorrefmark{2},
    Minh Tran\IEEEauthorrefmark{2},
    Jonathan Kua\IEEEauthorrefmark{1},
    Thuong Hoang\IEEEauthorrefmark{1} and
    Jiong Jin\IEEEauthorrefmark{3}
    } \\
    \IEEEauthorrefmark{1}Deakin University, Australia;
    \IEEEauthorrefmark{2}RMIT University, Vietnam;
    \IEEEauthorrefmark{3}Swinburne University of Technology, Australia; \\
    {\{peter.tran, jonathan.kua, thuong.hoang\}@deakin.edu.au}; {minh.tranquang@rmit.edu.vn}; {jiongjin@swin.edu.au} \\
}

\maketitle
\begin{abstract}
The persistent challenge in scaling authentic manipulator education within university laboratories is a structural dichotomy: commercial digital twins are often cost-prohibitive and rigidly scripted, whereas open-source robotics middleware (ROS) imposes steep technical and syntax barriers for novices. To resolve this logistical and educational friction, this paper proposes a scalable four-tier communication architecture tailored for sustainable robotic curricula. Rather than focusing on software application design, our study examines the underlying data exchange mechanisms required to bridge visual conceptual environments with physical robotic endpoints, utilizing the Graphical Open-Source Platform (GOSP) as a reference implementation. Our work details the framework's technical integration of 3D visual armature modeling with a robust ROS middleware backend, emphasizing the serialization, routing, and encapsulation of intricate communication routines. Preliminary sim-to-real validation using multi-axis spatial trajectories confirms that encapsulating these communication pipelines provides sufficient fidelity and a hardware-agnostic pathway. By bridging virtual design and physical execution, our architectural blueprint offers a viable infrastructure for engineering education.
\end{abstract}


\section{Introduction}
The integration of advanced robotic manipulators into industrial informatics curricula is imperative to meet the evolving demands of Industry 4.0 and industrial cyber-physical systems (ICPS)~\cite{cfa_tii_24,j_jin_cloud-fog_2025}. However, facilitating scalable hardware-in-the-loop training poses a formidable logistical and budgetary challenge for educational institutions. The physical deployment of industrial-grade robotics is systematically constrained by prohibitive capital expenditures, rigid safety protocols, and limited laboratory infrastructure~\cite{yarmand_pandemic_2025, iotvr_jamt_25}. Consequently, academic programs are frequently compelled to restrict the scope and frequency of iterative design-based learning (DBL). This persistent resource bottleneck inevitably compromises the depth of practical skill acquisition, generating a structural disparity in access to authentic engineering education.

This resource bottleneck is exacerbated by a structural dichotomy within contemporary simulation ecosystems, creating a congested learning load for entry-level engineers~\cite{zou2025synergy}. Proprietary industrial simulators (e.g., ABB RobotStudio, FANUC ROBOGUIDE, KUKA.Sim) deliver high-fidelity digital twins but enforce prohibitive licensing models and strict hardware dependencies~\cite{zhu_dt_2025}. Furthermore, to mitigate the risk of damaging physical assets, these commercial suites constrain users to rigid, procedural workflows, nullifying open-ended design exploration~\cite{li_au_2025, coltran_jii_25}. Conversely, open-source middleware frameworks, notably the Robot Operating System (ROS), democratize access and yield high task authenticity without financial barriers~\cite{quigley2009ros}. However, their inherently code-centric architecture necessitates the manual orchestration of complex software dependencies, build environments, and abstract Unified Robot Description Format (URDF) syntax. This steep operational overhead imposes a technical barrier, congesting the novice's processing capacity and detracting from the acquisition of fundamental kinematic principles.

To address this technical barrier, this work proposes a ``de-engineering'' educational pedagogy that functions as a scaffolding resolution. Rather than requiring novices to master industrial-grade software architecture simultaneously with fundamental robotics theory, this methodology systematically abstracts the system's operational friction. By providing a smart pedagogical scaffold, the approach isolates the formation of core concepts (coordinate frames, kinematic joints, and spatial reasoning) from underlying syntactic complexity~\cite{kapukotuwa_uniros_2025}. Serving as a conceptual bridge, this strategy mitigates the congested learning load. Consequently, it establishes a robust, hardware-agnostic foundation that facilitates a smooth transition toward subsequent advanced training on certified industrial platforms.

This paper proposes an integrated four-tier communication architecture designed for sustainable robotic curricula. Technically, our work explores the underlying data exchange mechanisms required to bridge visual conceptual environments with physical robotic endpoints, utilizing the pilot GOSP as a foundational instantiation. This paper details the technical integration of 3D visual armature modeling with a robust ROS middleware backend, emphasizing the serialization, routing, and encapsulation of intricate communication routines. Furthermore, by presenting preliminary sim-to-real validation of spatial trajectories, our work demonstrates that transparent, hardware-agnostic data pipelines can provide scalable, cost-effective infrastructure for advanced engineering education.

\begin{figure}[htbp]
\centerline{\includegraphics[width=\linewidth]{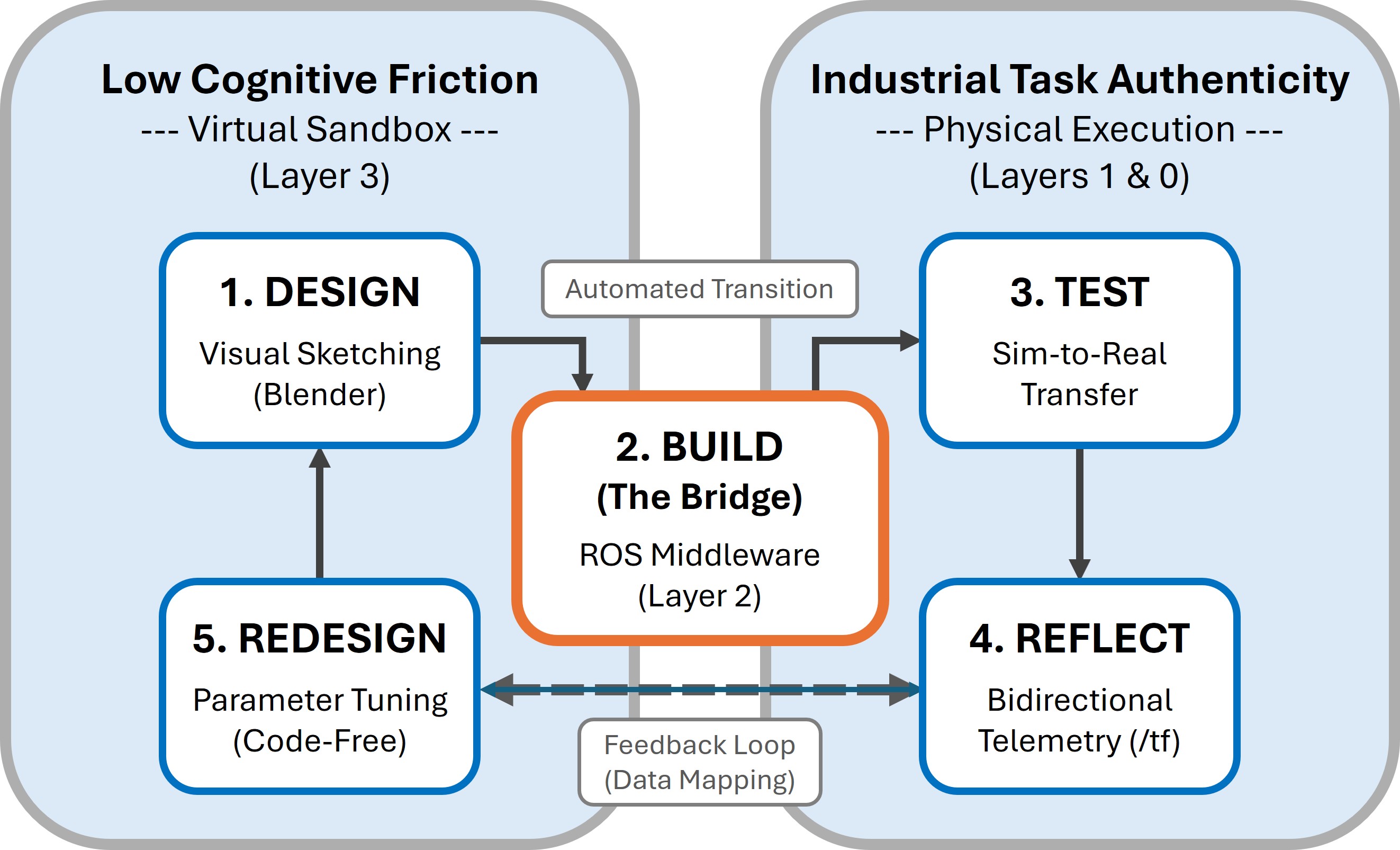}}
\caption{Structural alignment of the Graphical Open-Source Platform (GOSP) architecture with the Design-Based Learning (DBL) cycle. The platform's framework maps to the iterative learning phases: visual modeling (Layer 3) scaffolds the \lq\lq \textit{Design/Redesign}\rq\rq~phases, the automated Python middleware (Layer 2) facilitates the \lq\lq \textit{Build}\rq\rq~phase by abstracting operational friction, and the ROS-driven hardware endpoints (Layers 1 and 0) enable the \lq\lq \textit{Test/Reflect}\rq\rq~phases.}
\label{fig1}
\end{figure}

\section{Platform Philosophy and Pedagogical Alignment}
GOSP is engineered to function as an accessible, high-fidelity entry point for manipulator training, bridging visual design, dynamic simulation, and physical execution within a unified workflow. To address the structural dichotomy identified in contemporary tools, the platform's architecture is governed by two core engineering pillars that facilitate scaffolding pedagogy for robotic engineering education.

\subsection{Abstraction of Operational Friction}
To prevent novice processing capacity from being diverted toward troubleshooting fragmented open-source dependencies, GOSP systematically minimizes technical debt. By encapsulating complex control routines, socket communication, and ROS message serialization within a dedicated Python middleware layer, the system abstracts the operational friction. This allows learners to bypass boilerplate URDF syntax and maintain a primary focus on high-level systems engineering, coordinate frame manipulation, and fundamental kinematic algorithms.

\subsection{Safety-Driven Hardware Abstraction}
Industrial platforms inherently restrict iterative experimentation due to the safety risks and capital costs associated with physical manipulators. GOSP counteracts this by prioritizing a classroom-safe, virtual-first environment that faithfully mirrors industrial workflows. By decoupling the control logic from the physical hardware via ROS middleware, the platform provides a safe sandbox. Moreover, this hardware-agnostic architecture fosters exploratory design agency without the risk of catastrophic collisions on expensive laboratory assets.

\subsection{Structural Alignment with the DBL Cycle}\label{sec2c}
Effective DBL requires rapid iteration without the cognitive load of complex setups or restrictive licensing. GOSP eliminates these barriers, providing a cohesive, hardware-agnostic framework for a continuous learning cycle as shown in Fig.~\ref{fig1}:

\begin{itemize}
    \item \textbf{Design (Visual Scaffolding):} Learners construct kinematic chains using Blender's intuitive 3D armature environment. This visual-first approach allows novices to define joint constraints and spatial limits securely, anchoring their understanding in spatial reality.
    \item \textbf{Build (Automated Abstraction):} The platform automates middleware execution, translating visual models into operational ROS nodes. This encapsulation introduces students to modular system architecture and message passing while preventing early coding fatigue.
    \item \textbf{Test (Safe Experimentation):} Students conduct virtual verification for avoidance and joint-limit compliance within a risk-free digital sandbox. This is followed by a frictionless sim-to-real transition, deploying validated trajectories to hardware-agnostic educational endpoints.
    \item \textbf{Reflect (Knowledge Construction):} Live, bidirectional telemetry visually maps the physical robot's behavior back onto the digital twin. This real-time visual feedback loop empowers learners to analyze positional discrepancies and solidify robust mental models of 3D kinematics.
    \item \textbf{Redesign (Iterative Agency):} The platform enables rapid, code-free modifications of joint properties and trajectory parameters, allowing students to iterate on designs within a distraction-free ecosystem, transitioning from procedural troubleshooting to conceptual exploration.
\end{itemize}

\section{System Architecture}
\begin{figure*}[htbp]
\centerline{\includegraphics[width=0.923\linewidth]{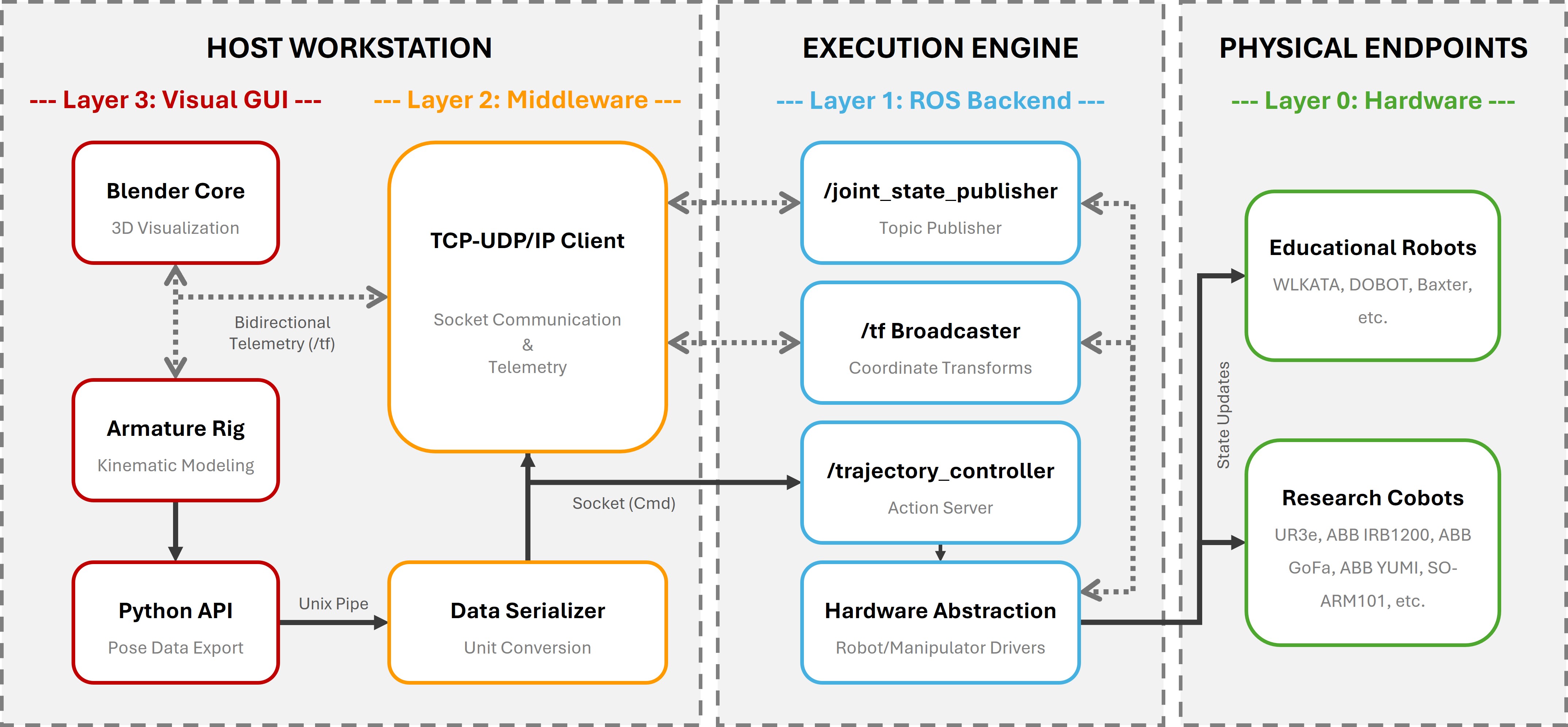}}
\caption{The four-tier communication architecture of the GOSP. A middleware bridge (L2) routes trajectory commands from the visual interface (L3) to the ROS execution engine (L1), establishing a closed-loop telemetry cycle with the hardware (L0).}
\label{fig2}
\end{figure*}

GOSP bridges visual design, simulation, and physical execution by encapsulating complex routines across four distinct integration layers, as illustrated in Fig.~\ref{fig2}.

\subsection{Layer 3: Graphical Interface (Blender)}
\begin{figure}[htbp]
\centerline{\includegraphics[width=0.75\linewidth]{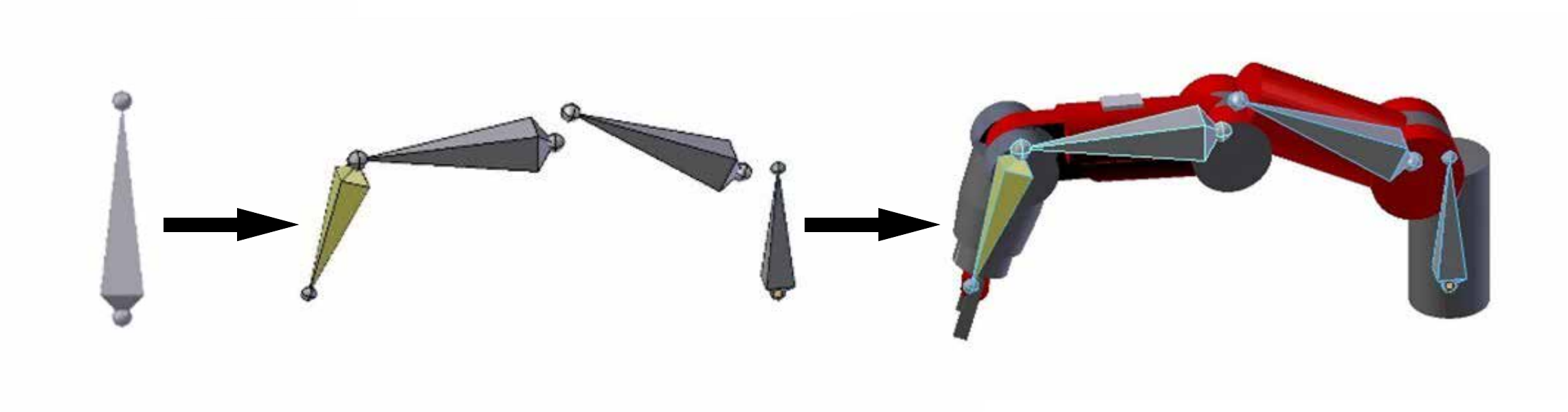}\includegraphics[width=0.28\linewidth]{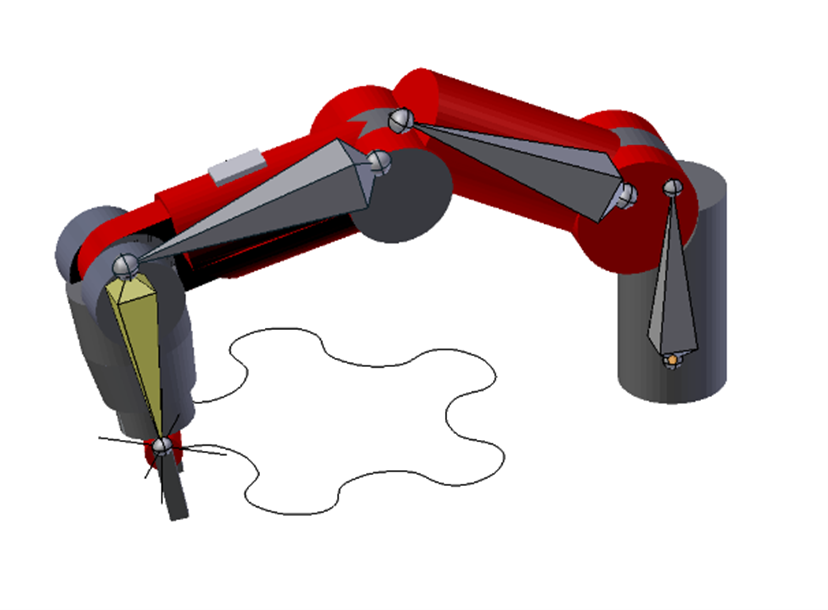}}
\caption{Kinematic mapping of a serial manipulator showcasing the armature skeleton in Blender and animated end-effector trajectories in Blender.}
\label{fig3}
\end{figure}

Serving as the primary user interface, the top layer leverages Blender to abstract the steep learning curve of URDF syntax in ROS. Learners construct manipulators utilizing Blender's armature system, a virtual skeleton comprising hierarchical joints and bones, as presented in Fig.~\ref{fig3}. Within this visual framework, bones represent rigid links, and joints serve as pivot points that define the kinematic chain of the robotic manipulator. The utilization of 3D environments to visually author and export complex kinematic structures has been proven to significantly reduce the URDF modeling bottleneck in advanced robotics research. This digital rig allows learners to intuitively assign degrees of freedom (DoFs), rotation limits, and motion paths. A native Python API then exposes these joint and frame parameters, exporting internal coordinate data to facilitate external control to the robot's digital twin or physical robotic manipulator.

\subsection{Layer 2: Python Middleware}
A lightweight Python client functions as the critical translation bridge between the visual environment and the execution layer. It serializes Blender's internal data, including poses, joint configurations, and motion paths, into standardized formats compatible with robotics middleware. This layer manages crucial unit conversions (translating Blender's internal scaling to standard SI units) and coordinate frame synchronization. By employing a publisher-subscriber architecture via Unix piping and TCP/IP sockets, it interfaces with ROS to establish a bidirectional stream. This routes real-time physical feedback back into Blender, enabling visualization of the digital twin.

\subsection{Layer 1: ROS Control and Computation}
The execution layer leverages ROS2 to provide a modular hardware-abstraction interface. This architecture allows inverse kinematics solvers, motion planners, and hardware drivers to be seamlessly swapped without modifying the higher-level graphical logic. The system state is dynamically maintained via core ROS nodes: \texttt{/joint\_state\_publisher} aggregates position data, \texttt{/trajectory\_controller} governs execution timing, and \texttt{/tf} (transform) broadcasters preserve the exact spatial relationships of the robot's links within 3D space.

\subsection{Layer 0: Hardware Execution}
The physical layer acts as the final execution endpoint. Because of the robust abstraction provided by the ROS middleware, the command structures generated in Layer 3 can operate on diverse robotic hardware. This capability extends from accessible educational platforms such as WLKATA and DOBOT to advanced research platforms such as Baxter, UR3e/7e/12e, SO--ARM101 (LeRobot), and cobots. The recent emergence of highly capable, open-source hardware ecosystems has made complex teleoperation and manipulation viable at a fraction of traditional industrial costs. This hardware-agnostic methodology guarantees a smooth sim-to-real transition and supports highly scalable, cost-effective deployment in educational laboratories before moving to industrial training.

\section{Preliminary Sim-to-Real Validation}
\begin{figure}[htbp]
\centerline{\includegraphics[width=.8\linewidth]{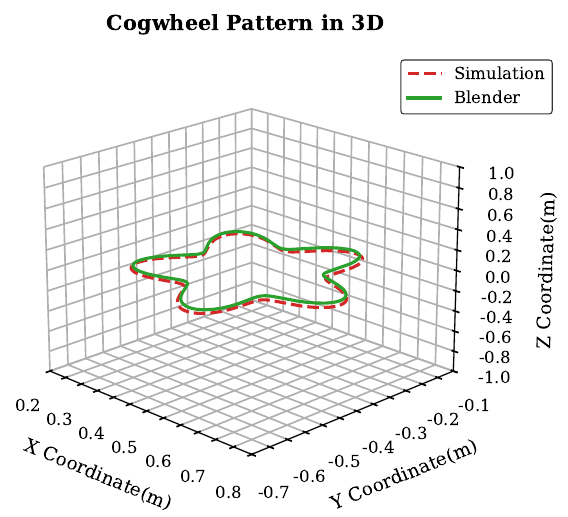}}
\centerline{\includegraphics[width=.9\linewidth]{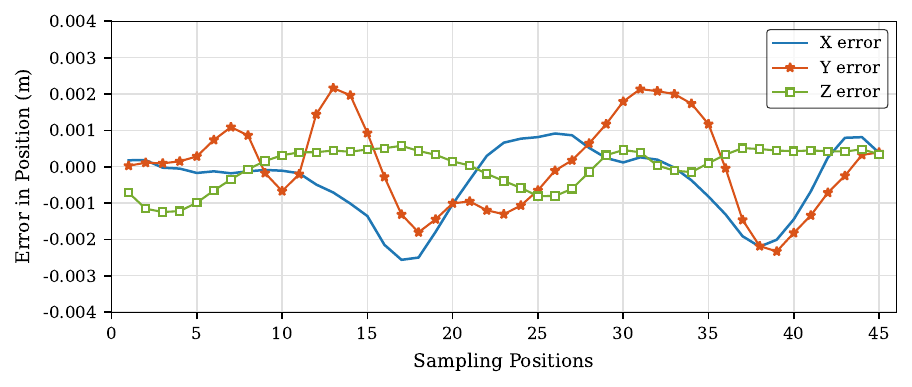}}
\caption{Sim-to-real validation of spatial trajectories. (Top) Planned graphical path (Blender) vs. executed simulation for a complex cogwheel pattern. (Bottom) Resulting positional errors (m) across axes, demonstrating high trajectory fidelity.} 
\label{fig4}
\end{figure}

A core feature of GOSP is to ensure that visual experimentation effectively translates into physical engineering competence. To validate the alignment between the virtual simulation and physical execution, empirical benchmark trajectory tests were conducted utilizing continuous, multi-axis spatial patterns. Specifically, the system was tasked with executing continuous paths, such as the multi-axis cogwheel pattern depicted in Fig.~\ref{fig4}. These trajectories were explicitly selected to rigorously test the middleware's synchronization capabilities, the accuracy of the inverse kinematics solver, and the system's ability to interpolate motion smoothly across multiple joints simultaneously. Before physical deployment, path feasibility and singularity avoidance were verified visually within the Blender digital twin, as indicated in Fig.~\ref{fig3}. This deliberate progression from risk-free virtual verification to authentic hardware execution directly mirrors the scaffolding pedagogy required for sustainable standardized industrial training.

\newpage
Quantitative analysis reveals that the absolute positional error is tightly bounded within approximately $\pm0.0025$ m ($\pm2.5$ mm) across all axes. Furthermore, vertical (Z-axis) tracking demonstrates exceptional stability, with the greatest variance confined to the planar (X and Y) axes during rapid, multi-joint interpolation. Crucially, these empirical results validate the framework's structural alignment with the DBL cycle outlined in Section~\ref{sec2c}. During the \lq\lq\textit{Design}\rq\rq~and~\lq\lq\textit{Build}\rq\rq~phases, the complex multi-axis cogwheel trajectory was authored entirely within the visual sandbox, abstracting the underlying ROS middleware without degrading system fidelity. The tight error boundaries ($\pm2.5$ mm) observed during physical execution validate the \lq\lq\textit{Test}\rq\rq~phase, proving that novice engineering students can safely rely on the digital twin's collision and kinematic predictions before physical hardware deployment. 

Furthermore, the absence of critical latency or coordinate misalignment ensures the integrity of the \lq\lq\textit{Reflect}\rq\rq~phase. Because the bidirectional telemetry is accurate, any positional discrepancies analyzed by the learner reflect authentic mechanical behaviors rather than artificial software latency. By guaranteeing this fidelity data mapping, the communication architecture empowers students to engage in the \lq\lq\textit{Redesign}\rq\rq~phase, making iterative spatial adjustments with absolute confidence in the system's execution. Ultimately, this verifies that the hardware-agnostic DBL cycle remains authentic, safe, and highly effective for the learner~\cite{coltran_jii_25, iotvr_jamt_25}.

\newpage
\section{Conclusions and Future Work}
This paper presented the scalable four-tier communication architecture designed to address the structural dichotomy and resource constraints currently limiting authentic manipulator training for robotic curriculum. By evaluating the system through the pilot GOSP instantiation, we validated that encapsulating distributed middleware telemetry and ROS data routing into the unified framework bridges virtual conceptual design with physical system execution. Preliminary sim-to-real validation demonstrated that the architecture maintains sub-centimeter trajectory fidelity, confirming its technical reliability for spatial operations without imposing restrictive licensing requirements or critical latency on engineering practices.

Having established the technical robustness, our ongoing research will transition toward empirical, human-centered evaluation. Future work will assess the pedagogical efficacy within active DBL frameworks, utilizing this pilot GOSP, quantifying how abstracting operational friction mitigates extraneous cognitive load and accelerates fundamental kinematic comprehension in novice robotic engineers. By evaluating these metrics before students' exposure to standardized commercial certification suites, our research ultimately aims to provide universities with a proven, sustainable, and cost-effective blueprint for deploying advanced engineering curricula.

\bibliographystyle{IEEEtran}
\bibliography{refs}

\end{document}